\documentclass[aps,prl,preprint,superscriptaddress]{revtex4}
\usepackage{graphicx}
\usepackage{verbatim}

\begin{document}

\title{Electron dynamics in topological insulator semiconductor-metal interfaces (topological p-n interface)}

\author{L. Andrew Wray}
\affiliation{Department of Physics, Joseph Henry Laboratories, Princeton University, Princeton, NJ 08544, USA}
\affiliation{Advanced Light Source, Lawrence Berkeley National Laboratory, Berkeley, California 94305, USA}
\author{Suyang Xu}
\author{Madhab Neupane}
\author{David Hsieh}
\affiliation{Department of Physics, Joseph Henry Laboratories, Princeton University, Princeton, NJ 08544, USA}
\author{Dong Qian}
\affiliation{Department of Physics, Joseph Henry Laboratories, Princeton University, Princeton, NJ 08544, USA}
\affiliation{Department of Physics, Shanghai Jiao Tong University, Shanghai 200030, China}
\author{Alexei V. Fedorov}
\affiliation{Advanced Light Source, Lawrence Berkeley National Laboratory, Berkeley, California 94305, USA}
\author{Hsin Lin}
\author{Susmita Basak}
\author{Arun Bansil}
\affiliation{Department of Physics, Northeastern University, Boston, MA 02115, USA}
\author{Yew San Hor}
\author{Robert J. Cava}
\affiliation{Department of Chemistry, Princeton University, Princeton, NJ 08544, USA}
\author{M. Zahid Hasan}
\affiliation{Department of Physics, Joseph Henry Laboratories, Princeton University, Princeton, NJ 08544, USA}



\date{\today}

\maketitle

\textbf{Single-Dirac-cone topological insulators (TI) are the first experimentally discovered class of three dimensional topologically ordered electronic systems, and feature robust, massless spin-helical conducting surface states that appear at any interface between a topological insulator and normal matter that lacks the topological insulator ordering \cite{Intro,TIbasic,DavidNat1,DavidScience,DavidTunable,MatthewNatPhys}. This topologically defined surface environment has been theoretically identified as a promising platform for observing a wide range of new physical phenomena, and possesses ideal properties for advanced electronics such as spin-polarized conductivity and suppressed scattering \cite{DavidScience,dhlee,MacDonaldKerr,FuNew,FerroSplitting,FuHexagonal,TopoFieldTheory,Biswas,ExcitCapacitor,palee,ZhangDyon,KaneDevice,DavidNat1,MatthewNatPhys,ZhangPred,DavidTunable,YazdaniBack,WrayCuBiSe,WrayFe,YeHelix}.
A key missing step in enabling these applications is to understand how topologically ordered electrons respond to the interfaces and surface structures that constitute a device. Here we explore this question by using the surface deposition of cathode (Cu/In/Fe) and anode materials (NO$_2$) and control of bulk doping in TI Bi$_2$Se$_3$ from P-type to N-type charge transport regimes to generate a range of topological insulator interface scenarios that are fundamental to device development. The interplay of conventional semiconductor junction physics and three dimensional topological electronic order is observed to generate novel junction behaviors that go beyond the doped-insulator paradigm of conventional semiconductor devices and greatly alter the known spin-orbit interface phenomenon of Rashba splitting. Our measurements for the first time reveal new classes of diode-like configurations that can create a gap in the interface electron density near a topological Dirac point and systematically modify the topological surface state Dirac velocity, allowing far reaching control of spin-textured helical Dirac electrons inside the interface and creating advantages for TI superconductors as a Majorana fermion platform over spin-orbit semiconductors.}

The surface state of topological insulator Bi$_2$Se$_3$ crystals cleaved in ultra high vacuum (UHV) is a singly degenerate Dirac cone that spans a bulk semiconductor band gap of $\Delta$$\gtrsim$300meV \cite{DavidTunable,MatthewNatPhys,ZhangPred,WrayCuBiSe}, with a Fermi level that can be easily moved by bulk dopants (Fig. 1a). These surface electron dynamics, identified in 2008, have inspired an ongoing stream of proposals anticipating new phenomena and technologies that can be achieved by interfacing conventional electronics with two dimensional spin-chiral Dirac electrons, including the possibility of observing exotic quasiparticles such as Majorana fermions, dyons and axions with properties that have been found nowhere else in the universe \cite{FuNew,ZhangDyon,TopoFieldTheory}. Developments from the last few months have demonstrated that it is possible to fabricate nanodevices using Bi$_2$Se$_3$ and pass current through the surface states \cite{topGate,sacebeDevice}, but have also highlighted the fact that topological insulators are a new state of matter distinct from normal metals and insulators, and their properties in a junction setting are not yet known. Changes in the surface chemistry of Bi$_2$Se$_3$ have been found to modify the Dirac cone kinetics in unpredictable ways \cite{WrayCuBiSe} and induce bulk electrons to be trapped in new surface Dirac cones \cite{WrayFe}. To bridge the gap between the conceptual promise of topological insulator Dirac electrons and real devices that can take advantage of topological electronic order, it is critical to develop an experimentally motivated understanding of how interfaces between normal metals and topological insulators will function.


To determine how junctions will change the topological electron kinetics, we have deposited thin films on the surface of carrier doped Bi$_2$Se$_3$ to generate a wide range of semiconductor-metal P-N and N-N interfaces, and used angle resolved photoemission spectroscopy (ARPES) to map changes in the topological surface electrons. The non-TI part of the interface is grown by evaporative deposition of strong charge donors (Cu, NO$_2$, Fe), achieving conventional diode interface voltages up to V$_d$$\sim$0.7eV from a sub-nanometer surface layer that is permeable to photoexcited electrons \cite{univCurve}. The cleaved Se$^{2-}$ plane of Bi$_2$Se$_3$ is maintained at low temperature (T$<$20$^o$K) to provide a chemically neutral and homogeneous surface for these adatoms, as discussed in the online Supplementary Information (SI). Deposition depth is presented in units of monolayers (ML) to approximately represent the number of deposited ions per hexagonal unit cell of the surface, based on rough linear assumptions outlined in the SI. Summing the Fermi surface area of all surface states gives the surface state electron density or Luttinger count, which we find grows more slowly than the nominal rate of charge deposition with Cu$^+$ or Fe (Fe$^{2+}$ or Fe$^{3+}$) ions by approximately one order of magnitude. After an estimated 2ML of Cu deposition the Fermi level ceases to move as more adatoms are deposited, and very few additional electrons can be added to the surface states (Fig. \ref{fig:PNfig}b). As shown in Fig. 1c, the topological surface state is dramatically changed by the voltage drop in a P-N inteface between hole doped (P-type) Bi$_2$Se$_3$ and copper, manifesting new Dirac cones and a gap in the surface density of states near the original (D0) Dirac point at 0.6-0.7eV binding energy.

The observation of a gap in the surface electron density is extremely surprising, because it is well known from theory that non-magnetic perturbations cannot open a true band gap in topological surface state Dirac points \cite{Intro, TIbasic}. To understand how the junction with non-magnetic copper could cause such a feature to appear, we have created a Green's function implementation of the experimentally-based \textbf{k}$\cdot$\textbf{p} model in Ref. \cite{FuNew,FuHexagonal} to numerically simulate junctions at the surface of a \emph{semi-infinite} topologically ordered Bi$_2$Se$_3$ slab based on experimentally measured bulk electron kinetics from Ref. \cite{WrayCuBiSe,fisherBending}. Unlike other current numerical models, this approach is capable of simulating the full screening depth of electrostatic interactions inside a Bi$_2$Se$_3$ junction ($\sim$100nm), and encompasses both the experimentally determined bulk kinetics and the underlying topological order of electrons in the simulated crystal. Using a screening potential that matches the observed change in topological surface state electron energies in the Bi$_2$Se$_3$-Cu P-N interface ($\Delta_\mu$=0.7eV), the simulation shows that the upper Dirac cone surface state electrons are still found in the first quintuple atomic layer of Bi$_2$Se$_3$, but the corresponding lower Dirac cone electrons are found more than 1nm deeper in the crystal, due to hybridization with other surface states derived from the perturbed bulk valence band (Fig. 1b). Because the lower Dirac cone electrons of the original surface state are deeper in the crystal than the $\sim$1nm photoemission penetration depth, photoemission measurements are expected to show a void or ``gap" near the D0 Dirac point, even though the three dimensional band structure is not gapped. This effect observed from our data and model is consistent with a recent theoretical proposal that interfacing TI surface states with a parasitic (degenerate) 3D metal can reshape the surface bands to take the appearance of a band gap \cite{parMetal}, a coexistence expected in many theoretically predicted topologically ordered metals such as half-Heusler compounds \cite{HeuslerDiscovery}.



The incremental evolution leading to the appearance of multiple new Dirac cones on Cu-deposited P-type Bi$_2$Se$_3$ is shown in Fig. \ref{fig:PNfig}. The build-up of positively charged surface ions first donates electrons to the original Dirac cone of freshly cleaved Bi$_2$Se$_3$, and then gradually binds four new surface states that intersect in two new Dirac points (D1 and D2) at the Brillouin zone (BZ) center. Tracing these new states from their points of intersection in the BZ center, we observe that they exchange partners and merge with a different surface band at momenta larger than $\sim$0.1$\AA^{-1}$, corresponding roughly to the outer contour of the region in which symmetries are inverted to bring about the TI state in numerical models. The resulting Fermi surface after heavy deposition (Fig. \ref{fig:PNfig}a, right) is made up of three rings, of which the outer two are doubly degenerate (within intrinsic width resolution). The doubly degenerate outer rings cannot carry a non-zero Berry's phase, meaning that the Berry's phase of $\pi$ by which a topological insulator interface is identified is obtained from the innermost singly degenerate Fermi surface ring, located deepest inside the Bi$_2$Se$_3$ crystal. The outermost ring is found at k$\sim$1.7-2.0$\AA^{-1}$, and is hexagonally warped into a doubly-degenerate ``snowflake" shape that resembles the singly degenerate hexagonal Fermi surface associated with magnetic instabilities in related Bi$_2$Te$_3$ \cite{FuHexagonal}. Although alternative terminologies have been proposed in recent literature \cite{HofmannRashba2DEG}, we refer to all of these surface bands as topological surface (or interface) states, because they manifest at the surface of the material and all five Dirac bands must be observed to measure the $\Theta_B$=$\pi$ topological Berry's phase.


In topologically trivial semiconductors with strong spin-orbit interactions, electrons near an interface are commonly spin-split into spin polarized Dirac cones by the Rashba effect. The partner swapping behavior we observe in the D1 and D2 Bi$_2$Se$_3$-Cu P-N interface states appears identical to the conventional Rashba effect in photoemission measurements at small momenta (K$\ll$0.1$\AA^{-1}$), but has very different phenomenological implications. The conventional Rashba effect causes surface electrons to be more strongly spin-polarized as their momentum deviates from the BZ center, due to the Rashba Hamiltonian term which is proportional to momentum ($\tilde{H}$$_R$=(2$\alpha_R$$\vec{k}\times\hat{z})\cdot\vec{s}$, with $\vec{s}$ representing spin). In contrast, though Bi$_2$Se$_3$-Cu P-N interface electrons obey this linear relation near their Dirac points, at large momentum the surface electrons are all nearly doubly-degenerate, and do not generate a large density of spin-polarized conduction states at the Fermi level. The difference can be seen clearly by comparing the Bi$_2$Se$_3$-Cu interface band structure with the surface states of BiTeI (Fig. \ref{fig:PNfig}c), a compound chemically similar to Bi$_2$Se$_3$, but which our measurements show to be topologically trivial and have an extremely strong Rashba effect ($\alpha_R$=2.2$\pm$0.1eV*$\AA$, see SI). Due to the partner swapping effect, spin-helical conduction states inside the interface are still limited to a single Dirac cone as is desired for applications.




The upper Dirac cone surface state of the N-type Bi$_2$Se$_3$ samples explored in this study (including Cu$_x$Bi$_2$Se$_3$) has a momentum-axis width ($\delta$k) of $\delta$k$\sim$0.035$\AA^{-1}$ at half maximum intensity before deposition, which can be inverted to reveal the electronic mean free path (mfp) of $\ell$=1/$\delta$k=29$\AA$. Momentum-axis width of the original Dirac cone increases steadily as Cu is added to the surface (Fig. \ref{fig:PNfig}d) and is surprisingly broad when one considers that disorder-induced scattering is suppressed by the spin-momentum locking of electrons \cite{DavidScience,DavidTunable,YazdaniBack}. Depositing magnetic iron ions instead of non-magnetic copper results in an unusual fluctuation in $\delta$k that has been discussed in Ref. \cite{WrayFe} and may be due to reduced disorder following a 2D ferromagnetic ordering transition. The new surface states generated by copper deposition have sharper momentum widths of $\sim$0.027$\pm$3$\AA^{-1}$ ($\ell$=37$\AA$ mfp) just above the 0.27eV binding energy D1 Dirac point in Fig. \ref{fig:PNfig}a(right) and $\sim$0.023$\pm$3$\AA^{-1}$ ($\ell$=43$\AA$ mfp) just above the 0.1eV D2 Dirac point. The capability to generate a longer mean free path of spin polarized surface electrons via deposition or top gating, as identified in Fig. \ref{fig:PNfig}, is desirable to enhance long range magnetic \cite{FuHexagonal,WrayFe,YeHelix} and non-Abelian coherent behaviors unique to topological insulators \cite{WrayCuBiSe,FuNew}.

Topological insulators are among the very few material systems with band structures that can be manipulated to achieve transport defined by a Dirac point \cite{DavidTunable}, and most proposed applications for TIs derive from the behavior of low density spin-helical Dirac surface electrons. In Fig. \ref{fig:DPfig}c we present two realizations of P-N interfaces in which the Bi$_2$Se$_3$ surface state Dirac point is brought to the Fermi level where it will shape transport properties. The surface chemical potential of the N-type sample shown at left (as-grown Bi$_2$Se$_3$ with [Se]$^{2+}$ defects) is lowered by depositing NO$_2$$^-$, and the P-type sample (Bi$_{1.995}$Ca$_{0.005}$Se$_3$) was modified through deposition of In$^+$. Our data show that lowering the surface chemical potential of an N-type sample to the Dirac point has the effect of dramatically increasing the Dirac velocity from v$_D$=1.55$\pm$0.08eV$\cdot\AA$ to 2.30$\pm$0.08eV$\cdot\AA$, observed from the lower Dirac cone band slope traced in Fig. \ref{fig:DPfig}a,c. The velocity of lower Dirac cone electrons changes rapidly immediately after deposition begins, but stabilizes after the surface chemical potential falls below the bulk conduction band minimum 0.17$\pm$0.02 eV above the Dirac point (Fig. \ref{fig:DPfig}b). When the chemical potential lies inside the bulk electronic gap as measured by surface-sensitive photoemission, the lower Dirac cone velocity has the same value of 2.3$\pm$0.1eV$\cdot\AA$ for both N-type and P-type samples (Fig. \ref{fig:DPfig}b,c) even though the they are exposed to surface ions with opposite charge. As the chemical potential at the surface of Bi$_2$Se$_3$ is bent into the bulk band gap, numerical simulations show that the region near the surface becomes depleted of bulk conduction electrons (Fig. \ref{fig:DPfig}d and Fig. S5 in the SI), causing bulk and surface charge carriers to no longer spatially overlap with one another. The bulk and surface charge carriers occupy the same atomic orbitals, and spatial overlap between the surface and bulk conduction electrons will result in strong on-site charge and spin interactions that are likely responsible for the tunable Dirac velocity observed in our data.



Interfaces with like carrier type (N-N or P-P) are also of great technological importance, because they create a less resistive electronic junction with minimal charge inhomogeneity. Measurements in Fig. \ref{fig:SCfig} explore the effect of light copper deposition on the surface of N-type Cu$_{0.02}$Bi$_2$Se$_3$ to observe how an N-type surface layer interfaces with the electron doped topological insulator. Excess surface-deposited Cu adatoms cause the electron energies near the surface of the TI to bend towards higher binding energy, as they would in an N-N Schottky diode. A new surface state is fully formed after 0.08ML of Cu deposition (Fig. \ref{fig:SCfig}a, measurements at $<$0.08ML coverage in SI), and remains doubly degenerate within experimental resolution until approximately 0.15ML of Cu deposition, at which point it becomes spin-split as seen from a deviation between the band energy at the Brillouin zone center and the band energy minimum (Fig. \ref{fig:SCfig}c).


Because TI surface states cross the bulk band gap and have different energetics than bulk electrons \cite{TIbasic,DavidScience}, when interfacing non-TI materials with a topological insulator, the surface states will trap charge carriers and attract or repel bulk electrons even when the bulk carrier doping of each material is identical. Our ARPES measurements show that prior to depositing ions on the vacuum-cleaved Cu$_{0.02}$Bi$_2$Se$_3$ surface, a negative charge is trapped by the surface state, and will be screened over a range of more than 10 nanometers as simulated from the Poisson equation (Fig. \ref{fig:SCfig}e). Assuming that each Cu atom enters a +1 valence state, depositing 0.08ML of copper causes the net charge at the crystal surface to change from negative to positive. Between 0 and 0.08ML (shaded in Fig. \ref{fig:SCfig}e) the electric field penetrating beyond the first few nanometers of the crystal surface is expected to be very small, allowing electrons to retain their bulk-derived energetics and symmetries as they approach the surface from deeper in the crystal. As Cu deposition is raised to 0.16ML, electron density in the surface state grows more slowly than the rate of deposition, causing an increasing proportion of the surface charge to be screened by bulk electrons. Titrating with positive ions or a gating voltage ($\sim$0.1eV from our data) to achieve an electrically neutral surface is therefore a promising method to enhance bulk interactions with the surface states, such as magnetic and superconducting proximity effects.

Recent measurements have identified that the bulk doped TI Cu$_x$Bi$_2$Se$_3$ (x$\geq$0.1) is a type-2 N-type bulk superconductor (Ref. \cite{WrayCuBiSe} and Fig. 1a, left), and that magnetic vortices at the surface are likely to bind non-Abelian particles known as Majorana fermions that have not yet been observed in nature \cite{WrayCuBiSe,FuNew}. Our calculations in Fig. \ref{fig:SCfig}e show that an electric field from the negatively charged surface state will cause rapid phase fluctuations that suppress superconductivity if the surface charge is not neutralized (Fig. \ref{fig:SCfig}e, inset). Moreover, because vortices are repelled by regions of high superfluid density, this surface environment provides a new mechanism to control the dynamics of Majorana fermions and potentially observe their non-Abelian properties. For normal superconductors, surface perturbations suppress superfluid density, generating an attractive force that pins vortices in place \cite{VortexPin,VortexPush}. For a topological insulator however, our data show that adjusting the surface chemical energy to cancel out the negative surface state charge can enhance superfluid density under a probe, causing surface vortices to be repelled by the probe rather than pinned to it (Fig. \ref{fig:SCfig}d).

The control and exclusion of vortices is a subject of active research to improve the properties of superconductors \cite{VortexPush,KaneDevice}, and the intrinsically charged TI surface provides a new and promising platform for vortex manipulation. Taken collectively, the P-N and N-N interface electronic structures observed here reveal how interface effects such as Rashba spin-splitting and the superconducting proximity effect are changed by the presence of topological order, and how topological insulator interfaces can be tuned to adjust the average speed of surface Dirac electrons and interact with surface Majorana fermion vortices.

\textbf{Methods summary:}
Angle resolved photoemission spectroscopy (ARPES) measurements were performed at the Advanced Light Source beamlines 10 and 12 using 35.5-48 eV photons with better than 15 meV energy resolution and overall angular resolution better than 1$\%$ of the Brillouin zone (BZ). Samples were cleaved and measured at 15$^o$K, in a vacuum maintained below 8$\times$10$^{-11}$ Torr. Fe atoms were deposited using an e-beam heated evaporator at a rate of approximately 0.1$\AA$/minute, and Cu atoms were deposited at a rate of 0.4$\AA$/minute. A quartz micro-balance supplied by Leybold-Inficon with sub-Angstrom sensitivity was used to calibrate the deposition flow rate, and a factor of 2.5$\AA$/ML was used to estimate monolayer depth. Surface charge is estimated from the expected +1 ionization state of copper on the cleaved [111] Se surface \cite{WrayCuBiSe}. Adsorption of NO$_2$ molecules on Bi$_2$Se$_3$ was achieved by controlled \emph{in situ} exposures under static flow mode, with care to minimize photon exposure of the adsorbed surface. Crystal growth of large single crystals of Ca$_x$Bi$_{2-x}$Se$_3$ and Cu$_x$Bi$_2$Se$_3$ and numerical techniques used to simulate the perturbed TI surface are described in the SI.

\newpage

\begin{figure*}[t]
\includegraphics[width = 16cm]{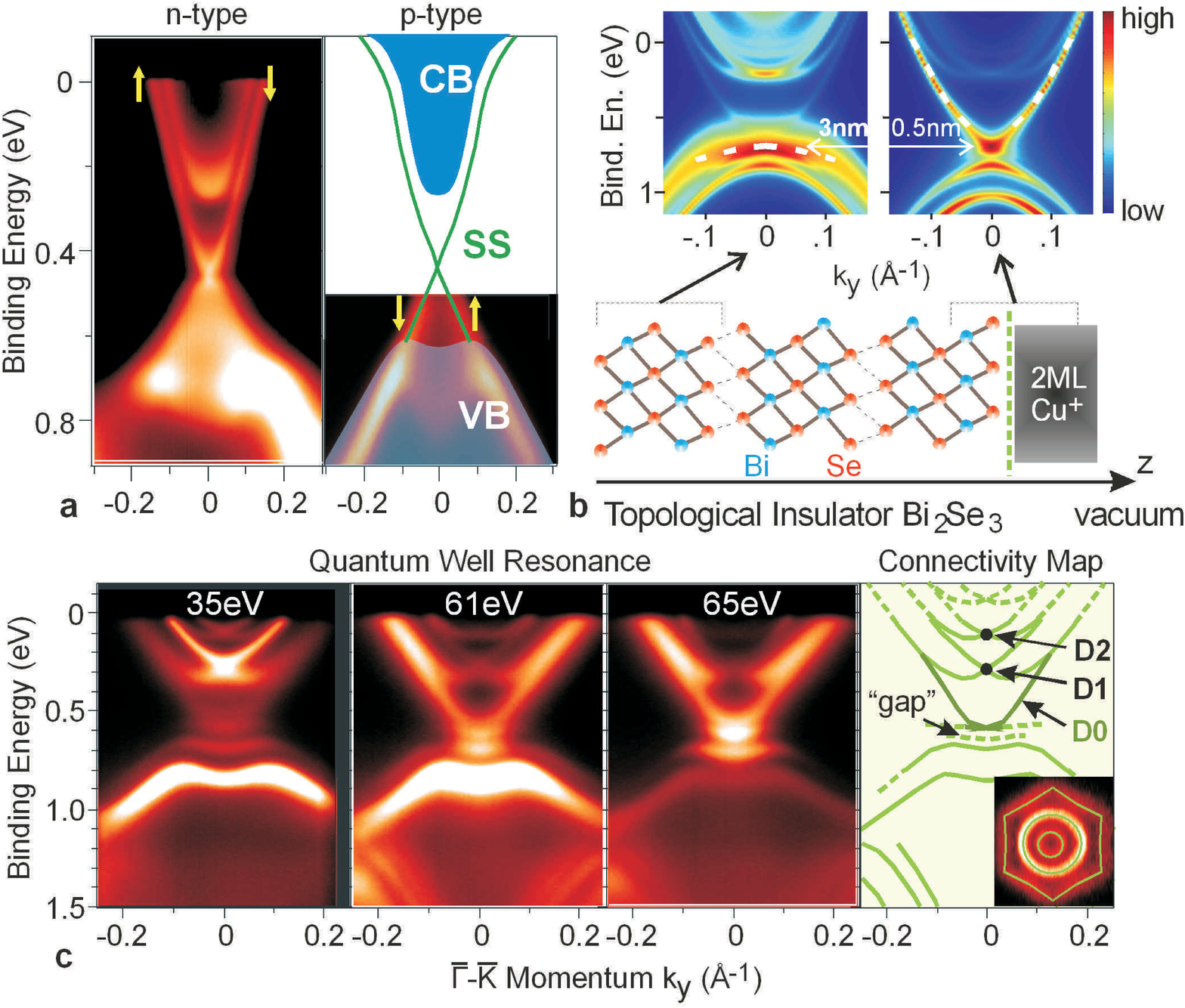}
\caption{{\bf{Coulomb-biased topological surface states}}: \textbf{a}, The (blue) bulk and (green) surface band structure of Bi$_2$Se$_3$ is traced above ARPES measurements of band structure at n-type and p-type bulk compositions. \textbf{b}, A numerical simulation of copper-interfaced Bi$_2$Se$_3$ shows the partial density of low energy surface states at depths of 0-0.5nm and 2-3nm inside the crystal. (white dashed lines) Electrons from the upper Dirac cone of Bi$_2$Se$_3$ remain in the outermost 0.5nm, and electrons in the lower Dirac cone are located 1-2nm deeper in the crystal. \textbf{c}, The surface states of Cu-deposited Bi$_2$Se$_3$ resonate at different photon energies, demonstrating that they occupy a quantum well-like environment. Measurements have a $\lesssim$1nm penetration depth into the crystal \cite{univCurve}. An integrated comparison reveals the band structure traced at right, with new Dirac points labeled D1 and D2 and the Fermi surface shown as an inset. States that evolve adiabatically from the original upper Dirac cone (``D0") are drawn in a darker hue.}
\end{figure*}

\begin{figure*}[t]
\includegraphics[width = 16cm]{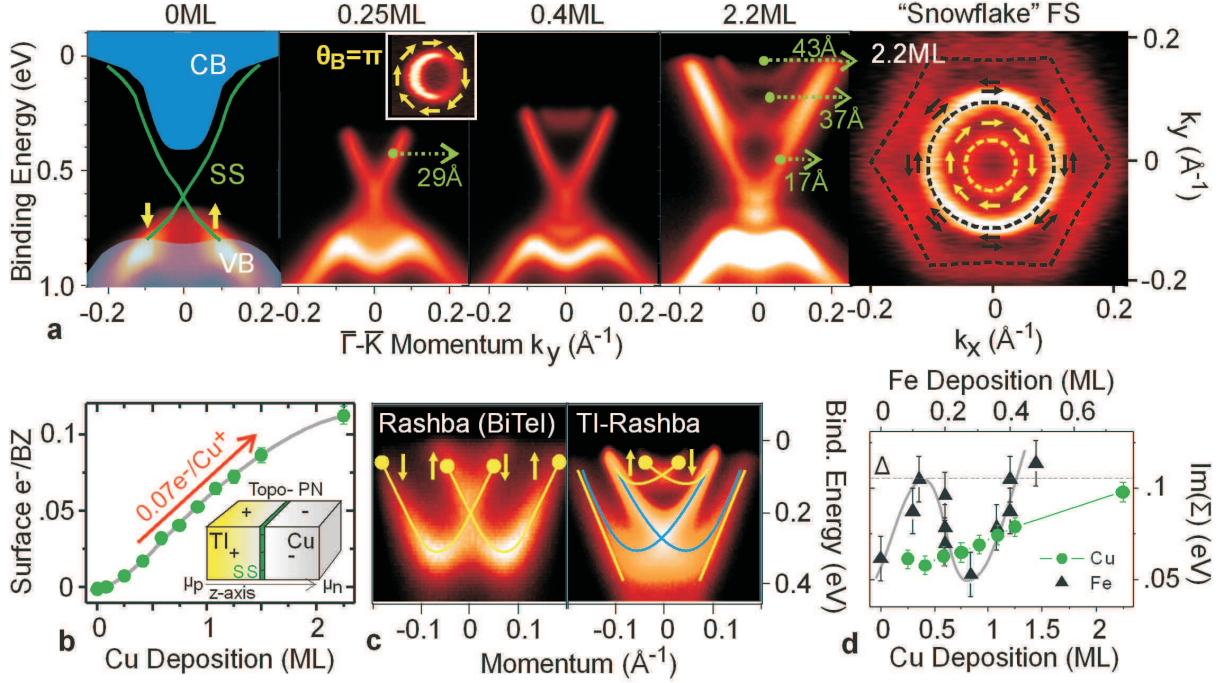}
\caption{\label{fig:PNfig}{\bf{A topological p-n interface}}: \textbf{a}, Surface deposition of copper is shown on bulk p-type Bi$_2$Se$_3$ (0.5$\%$ Ca doped as Bi$_{1.995}$Ca$_{0.005}$Se$_3$), with approximate pre-deposition electronic structure traced on the first panel and energies shifted to align surface band features between panels. The mean free path estimated from inverse momentum width is labeled in nanometers for surface conduction bands. Fermi surfaces are shown for (inset) 0.25ML and (right) 2.2ML of Cu deposition, with numerically predicted spin orientations labeled with arrows. \textbf{b}, The surface state carrier density estimated from photoemission data (Luttinger count) is plotted as a function of Cu deposition. \textbf{c}, (top) Large Rashba splitting in the surface states of topologically trivial BiTeI is compared with (bottom) the strikingly different Rashba-like effect in topologically ordered surface electrons. \textbf{d}, The imaginary self energy of the original (`D0') upper Dirac cone after Cu and Fe deposition is obtained by multiplying the momentum-axis width ($\delta$k) by half of the surface state velocity (Im($\Sigma$)=$\frac{\delta k \times3.5 eV-\AA}{2}$). Deposition depth axes are mismatched in proportion to the expected surface valence states of Cu$^+$ and Fe$^{3+}$.}
\end{figure*}

\begin{figure*}[t]
\includegraphics[width = 16cm]{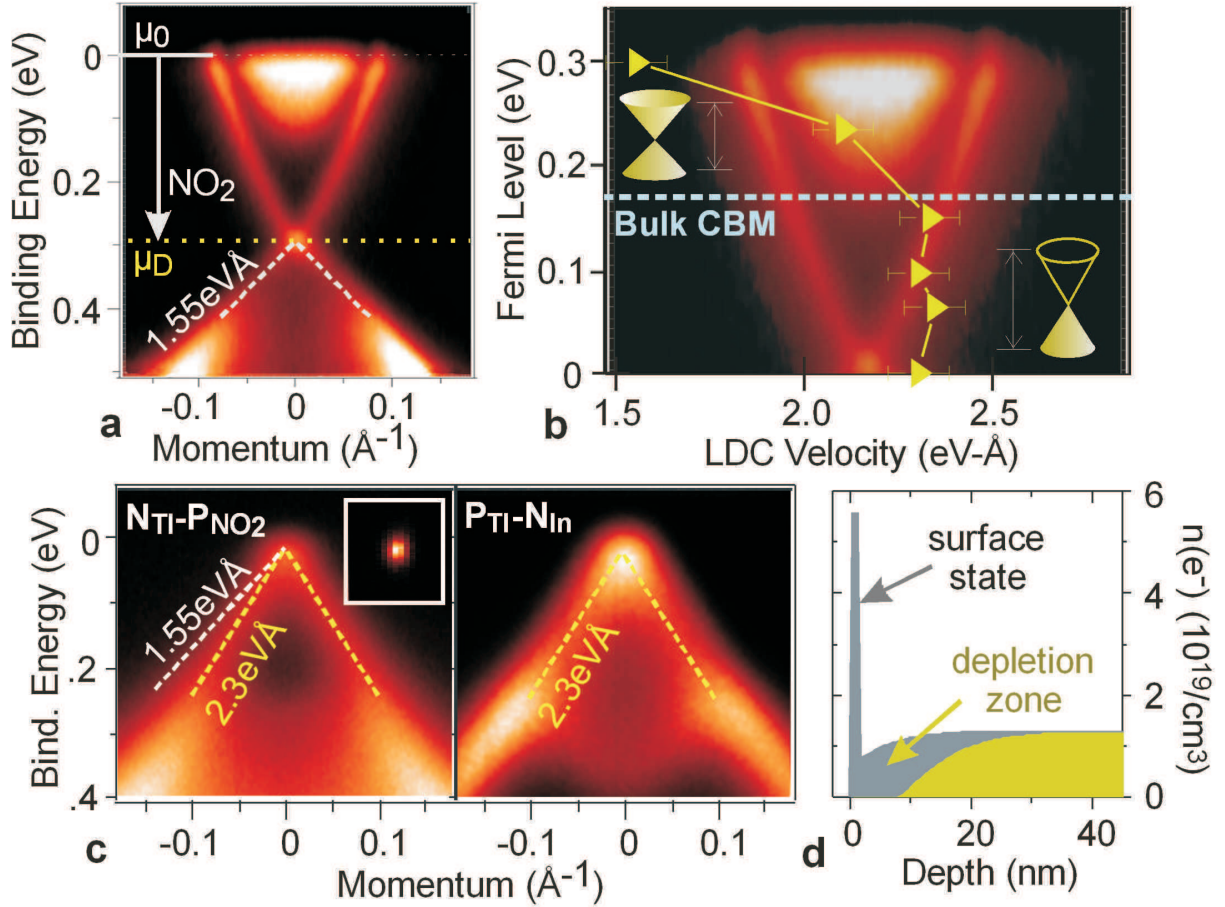}
\caption{\label{fig:DPfig}{\bf{Dirac point P-N interfaces}}: \textbf{a}, A measurement of band structure in N-type as-grown Bi$_2$Se$_3$ is labeled with the lower Dirac cone velocity and Dirac point energy ($\mu_D$) before surface deposition of NO$_2$. (\textbf{b}) The lower Dirac cone band velocity is plotted as a function of chemical potential above the Dirac point as NO$_2$ is added to the surface of N-type Bi$_2$Se$_3$. \textbf{c}, Photoemission images of bulk P-type and N-type samples with chemical potential set at the Dirac point via surface deposition of negatively and positively charged molecules, respectively (NO$_2$ and In). An inset shows the Dirac node Fermi surface, and the lower Dirac cone velocities (slopes) of the pristine pre-deposition surface states are traced in white. \textbf{d}, The electron density distribution is modeled for (gray) as-grown Bi$_2$Se$_3$ and (gold) the same sample with chemical potential set at the Dirac point by NO$_2$ deposition.}
\end{figure*}

\begin{figure*}[t]
\includegraphics[width = 16cm]{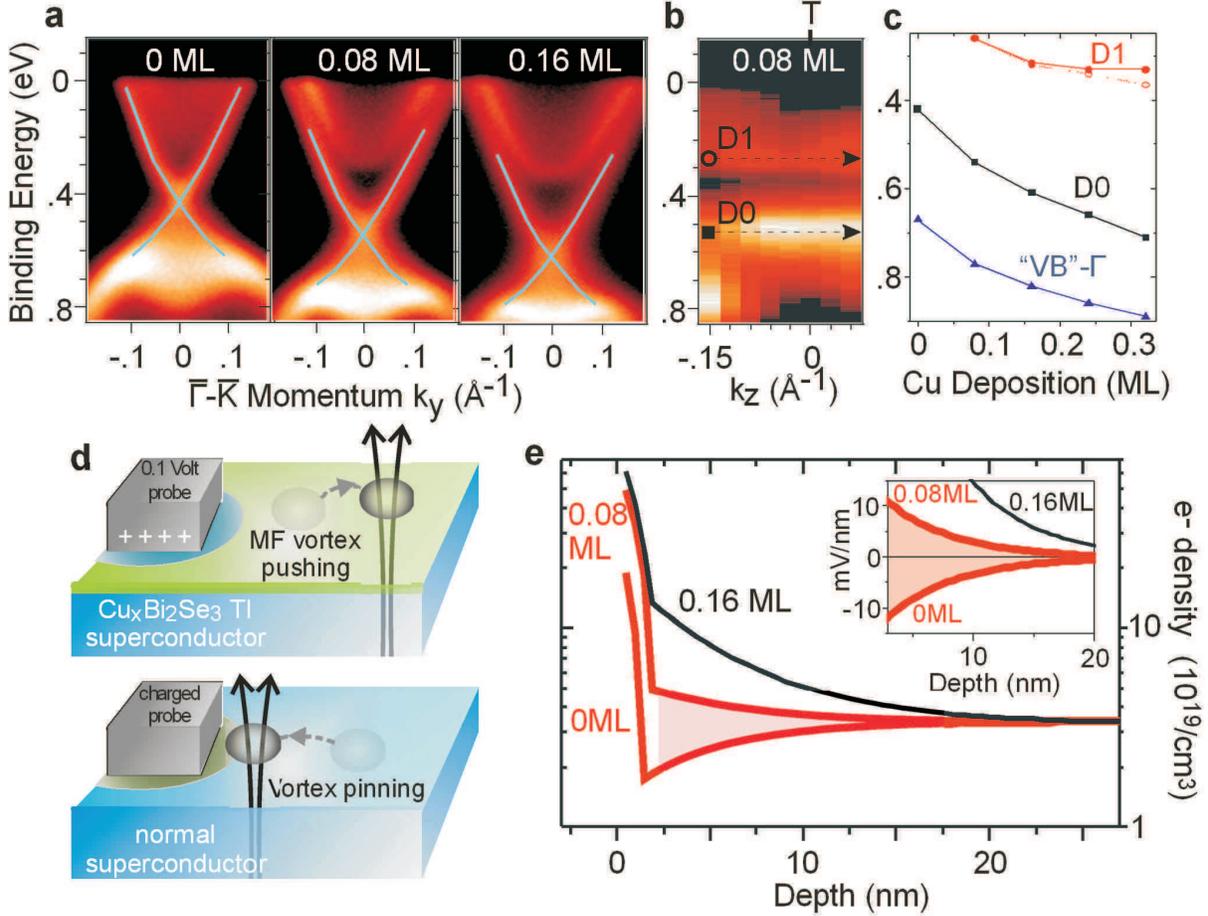}
\caption{\label{fig:SCfig}{\bf{Superconductors and n-n junctions}}: \textbf{a}, Photoemission measurements show the evolution of surface electron kinetics with light Cu surface deposition on bulk-doped Cu$_x$Bi$_2$Se$_3$. \textbf{b}, The new band that appears after 0.08 ML Cu deposition, labeled `D1' at the BZ center, has no momentum dispersion along the z-axis. The original surface state Dirac point is labeled `D0'. \textbf{c}, The doping evolution of band energies at the $\Gamma$-point is shown. The D1 band minimum, labeled `D1-min', and Dirac point energy (`D1') diverge after heavy deposition. \textbf{d}, Probes interfacing with normal superconductors suppress the superfluid density, attracting (``pinning") vortices. On a topological insulator, a gentle probe with bias voltage similar to 0.1eV can attract vortices by enhancing the superfluid density. Regions with suppressed superfluid density are indicated with green shading. \textbf{e}, The z-axis electron density distribution is modeled for different degrees of surface doping. (shaded region) Deposition levels between 0ML and 0.08 ML will reduce bulk screening, strengthening superconductivity near the surface. An inset shows the electric field inside the crystal.}
\end{figure*}

\end{document}